\documentclass[pre,aps,preprint,showpacs]{revtex4}
\usepackage{graphicx}
\usepackage{amsfonts}
\usepackage{amssymb}
\usepackage{amsmath}
\usepackage{textcomp}
\usepackage{enumerate}

\usepackage{color}
\usepackage{psfrag}

\usepackage{epsfig}

\begin{document}

\title{Avian photoreceptor patterns represent a disordered hyperuniform solution to a multiscale packing problem}

%\author{Yang Jiao,$^{1}$ and Salvatore Torquato$^{1,2}$}
%\email{torquato@princeton.edu} \affiliation{$^{1}$Princeton
%Institute of the Science and Technology of Materials,
%$^{2}$Department of Chemistry and Physics, Princeton Center for
%Theoretical Science, Princeton University, Princeton, New Jersey
%08544, USA}

\author{Yang Jiao$^{1, 2}$, Timothy Lau$^{3}$, Haralampos Hatzikirou$^{4,5}$, Michael
Meyer-Hermann$^{4}$, Joseph C. Corbo$^{3}$ and Salvatore
Torquato$^{1, 6}$} \affiliation{$^{1}$Princeton Institute of the
Science and Technology of Materials, Princeton University,
Princeton, NJ} \affiliation{$^{2}$Materials Science and
Engineering, Arizona State University, Tempe, AZ}
\affiliation{$^{3}$Department of Pathology and Immunology,
Washington University School of Medicine, St. Louis, MO}
\affiliation{$^{4}$Department of Systems Immunology, Helmholtz
Centre for Infection Research, Braunschweig, Germany}
\affiliation{$^{5}$Center for Advancing Electronics Dresden, TU
Dresden, 01062, Dresden, Germany} \affiliation{$^{6}$Department of
Chemistry, Department of Physics, Program in Computational and
Applied Mathematics, Princeton University, Princeton, NJ}

%\author{Yang Jiao}
%\affiliation{Physical Science in Oncology Center, Princeton
%University, Princeton New Jersey 08544, USA}

%\author{Timothy Lau}
%\affiliation{Department of Pathology and Immunology, Washington
%University School of Medicine, St. Louis, MO}

%\author{Haralampos Hatzikirou}
%\affiliation{Department of Systems Immunology, Helmholtz Centre
%for Infection Research, Braunschweig, Germany}

%\author{Michael Meyer-Hermann}
%\affiliation{Department of Pathology and
%Immunology, Washington University School of Medicine, St. Louis,
%MO}

%\author{Joseph C. Corbo}
%\email{jcorbo@pathology.wustl.edu} \affiliation{Department of
%Pathology and Immunology, Washington University School of
%Medicine, St. Louis, MO}

%\author{Salvatore Torquato}
%\email{torquato@electron.princeton.edu} \affiliation{Department of
%Chemistry, Princeton University, Princeton New Jersey 08544, USA}
%\affiliation{Department of Physics, Princeton University,
%Princeton New Jersey 08544, USA} \affiliation{Physical Science in
%Oncology Center, Princeton University, Princeton New Jersey 08544,
%USA} \affiliation{Program in Applied and Computational
%Mathematics, Princeton University, Princeton New Jersey 08544,
%USA}

\date{\today}


%\begin{article}

\begin{abstract}

Optimal spatial sampling of light rigorously requires that
identical photoreceptors be arranged in perfectly regular arrays
in two dimensions. Examples of such perfect arrays in nature
include the compound eyes of insects and the nearly crystalline
photoreceptor patterns of some fish and reptiles. Birds are highly
visual animals with five different cone photoreceptor subtypes,
yet their photoreceptor patterns are not perfectly regular. By
analyzing the chicken cone photoreceptor system consisting of five
different cell types using a variety of sensitive microstructural
descriptors, we find that the disordered photoreceptor patterns
are ``hyperuniform'' (exhibiting vanishing infinite-wavelength
density fluctuations), a property that had heretofore been
identified in a unique subset of physical systems, but had never
been observed in any living organism. A disordered hyperuniform
many-body system is an exotic state of matter that behaves like a
perfect crystal or quasicrystal in the manner in which it
suppresses large-scale density fluctuations and yet, like a liquid
or glass, is statistically isotropic with no Bragg peaks.
Remarkably, the photoreceptor patterns of both the total
population and the individual cell types are simultaneously
hyperuniform. We term such patterns ``multi-hyperuniform'' because
multiple distinct subsets of the overall point pattern are
themselves hyperuniform. We have devised a unique multiscale cell
packing model in two dimensions that suggests that photoreceptor
types interact with both short- and long-ranged repulsive forces
and that the resultant competition between the types gives rise to
the aforementioned singular spatial features characterizing the
system, including multi-hyperuniformity. These findings suggest
that a disordered hyperuniform pattern may represent the most
uniform sampling arrangement attainable in the avian system, given
intrinsic packing constraints within the photoreceptor epithelium.
In addition, they show how fundamental physical constraints can
change the course of a biological optimization process. Our
results suggest that multi-hyperuniform disordered structures have
implications for the design of materials with novel physical
properties and therefore may represent a fruitful area for future
research.

\end{abstract}

%\abbreviations{SAM, self-assembled monolayer; OTS, octadecyltrichlorosilane}

\maketitle

\section{Introduction}

The purpose of a visual system is to sample light in such a way as
to provide an animal with actionable knowledge of its surroundings
that will permit it to survive and reproduce \cite{ref19}. In most
cases, this goal is achieved most effectively by a highly regular
two-dimensional (2D) array of photoreceptors that evenly sample
incoming light to produce an accurate representation of the visual
scene. Classical sampling theory \cite{ref1, ref20} as well as
more recent studies \cite{ref2, ref3, ref21} have demonstrated
that the optimal arrangement of a 2D array of detectors is a
triangular lattice (i.e., a hexagonal array). Indeed, modeling
studies suggest that any deviation from a perfectly regular
arrangement of photoreceptors will cause deterioration in the
quality of the image produced by a retina \cite{ref4}.
Accordingly, many species have evolved an optimal sampling
arrangement of their photoreceptors. For example, the insect
compound eye consists of a perfect hexagonal array of
photoreceptive ommatidia \cite{ref5, ref6}. In addition, many
teleost fish \cite{ref7, ref8, ref9} and some reptiles
\cite{ref10} possess nearly crystalline arrangements of
photoreceptors. These and other examples attest that a perfect or
nearly perfectly ordered arrangement of photoreceptors can be
realized in a biological system.

\begin{figure*}
\begin{center}
\includegraphics[width=12.5cm,keepaspectratio]{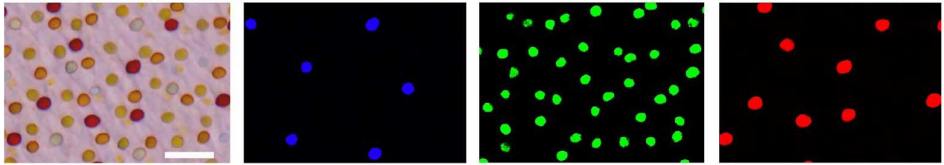}\\
\end{center}
\caption{Spatial arrangements of chicken cone photoreceptors. The
leftmost panel shows a flatmount preparation of a post-hatch day
15 chicken retina with colored oil droplets within the inner
segments of the five cone photoreceptor types. Size bar = 10
$\mu$m. The additional panels (from left to right) depict the same
field of view under illumination with ultraviolet, blue and green
light, respectively. Oil droplet auto-fluorescence permits
sub-type classification of the individual photoreceptor cells and
determination of their spatial coordinates.  These figures derive
from Ref. \cite{ref12}.} \label{fig1_cone}
\end{figure*}

Diurnal birds have one of the most sophisticated cone visual
systems of any vertebrate, consisting of four types of single cone
(violet, blue, green and red) which mediate color vision and
double cones involved in luminance detection \cite{ref22, ref23,
ref24}; see Fig. \ref{fig1_cone}. Despite the presence of numerous
evolutionary specializations in the avian eye, the overall
arrangement of bird cone photoreceptors is not perfectly ordered
but rather is irregular \cite{ref11, ref12}. The five avian cone
types exist as five independent, spatial patterns, all embedded
within a single monolayered epithelium. The individual cone
patterns in the bird's retina are arranged such that cones of one
type almost never occur in the near vicinity of other cones of the
same type \cite{ref12}. In this way, the bird achieves a much more
uniform arrangement of each of the cone types than would exist in
a random (Poisson) pattern of points.

Given the utility of the perfect triangular-lattice arrangement of
photoreceptors for vision \cite{ref4}, the presence of disorder in
the spatial arrangement of avian cone patterns is puzzling. It is
crucial to ascertain whether the apparent ``disordered''
photoreceptor arrangements correspond to a different optimal
solution because of constraints, such as cell size polydispersity,
that are not present in the aforementioned insect retinas. By
analyzing the chicken cone photoreceptor system using a variety of
sensitive microstructural descriptors that arise in statistical
mechanics and particle-packing theory \cite{ref27}, we show here
that the avian system possesses a remarkable type of correlated
disorder at large length scales known as hyperuniformity
\cite{ref13}, which has heretofore not been observed in a living
organism. Moreover, in a departure from any known physical system,
the patterns of both the total population and the individual cell
types are simultaneously hyperuniform, which we term {\it
multi-hyperuniformity}. We subsequently determine effective
interactions between the photoreceptors on multiple length scales
that could possibly explain their unusual disordered hyperuniform
state. Specifically, we consider two types of interactions that
have competing effects in determining the degree of order in the
system. Indeed, we show via computer simulations that the
local-energy minimizing configurations of such a many-particle
interacting system quantitatively capture, with high accuracy, the
unique spatial characteristics, including multi-hyperuniformity.
The fact that the aforementioned competing interactions lead to
disordered hyperuniform systems suggests that the photoreceptor
patterns may represent the most uniform sampling arrangement
attainable in the avian system due to intrinsic packing
constraints associated with the photoreceptor cells.

%Moreover, our findings also show how fundamental physical
%constraints could be acting to change the course of a biological
%optimization process. However, an evolutionary explanation for the
%development of the avian photoreceptor system is beyond the scope
%of this investigation.

The rest of the paper is organized as follows: In Sec. II, we
provide definitions of fundamental concepts used in our analysis
including various statistical microstructural descriptors, order
metrics as well as the concept of hyperuniformity. In Sec. III, we
quantitatively investigate structural characteristics of avian
photoreceptor patterns containing multiple cell species and show
that both the overall pattern and the arrangements of individual
species are hyperuniform. In Sec. IV, we determine the effective
interactions between the photoreceptors and devise a unique
multiscale packing model incorporating such interactions. We show
that our multiscale packing model can lead to point configurations
that are virtually indistinguishable from the actual photoreceptor
arrangements. In Sec. V, we provide concluding remarks.

\section{Definitions and fundamental concepts}

Before presenting our analysis of the avian photoreceptor system,
we first briefly review the ``hyperuniformity'' concept and its
quantification, which plays a central role in this paper. In
addition, we introduce the order metrics that will be employed to
characterize the avian patterns.

\subsection{Hyperuniform Systems}

The ensemble-averaged structure factor of infinite point configurations
in $d$-dimensional Euclidean space at number density $\rho$ is defined via
\begin{equation}
\label{eq_Sk} S({\bf k})= 1+ \rho {\tilde h}({\bf k}),
\end{equation}
where ${\tilde h}({\bf k})$ is the Fourier transform of the total
correlation function $h({\bf r}) = g_2({\bf r})-1$ and $g_2({\bf
r})$ is the pair correlation function of the system. Note that definition (\ref{eq_Sk}) implies  that the forward
scattering contribution to the diffraction pattern is omitted. For statistically homogeneous
and isotropic systems, the focus of this paper, $g_2$ depends on
the radial distance $r \equiv |{\bf r}|$ between the points (cell centers) as well
as the number density $\rho$. In two dimensions, the quantity $\rho g_2(r)2\pi r dr$
is proportional to the conditional probability of finding a cell
center at a distance between $r$ and $r+dr$ given that a cell
center is at the origin, where $\rho$ is
the number of cell centers per unit area.

The small-$k$ behavior of the structure factor $S(k)$ encodes information
about large-scale spatial correlations in the system and in the limit
$k \rightarrow 0$ determines whether the system is hyperuniform.
Specifically, an infinite point configuration in $d$-dimensional Euclidean space
is {\it hyperuniform} if
\begin{equation}
\label{eq_Sk2} \lim_{k\rightarrow 0}S(k) = 0,
\end{equation}
which implies that the infinite-wavelength density fluctuations of the system
(when appropriately scaled) vanish \cite{ref13}.

For computational purposes, the structure factor $S({\bf k})$ for
a given finite point configuration can be obtained directly from
the positions of the points ${\bf r}_j$ \cite{chase}, i.e.,
\begin{equation}
S({\bf k}) = \frac{1}{N} \left |{\sum_{j=1}^N \exp(i {\bf k} \cdot
{\bf r}_j)}\right |^2 \quad ({\bf k} \neq {\bf 0}),
\label{coll}
\end{equation}
where $N$ is the total number points in the system (under periodic boundary
conditions) and ${\bf k}$
is the wave vector. Note that the forward scattering contribution (${\bf k} = 0$)
in (\ref{coll}) is
omitted, which makes relation (\ref{coll})  completely consistent with the definition
(\ref{eq_Sk}) in the ergodic infinite-system limit. For statistically homogeneous
and isotropic systems, the focus of this
paper, the structure factor $S(k)$ only depends on the magnitude
of the scalar wavenumber $k = |{\bf k}| = 2\pi n/L$, where $n = 0,
1, 2\ldots$ and $L$ is the linear size of the system.

A hyperuniform point configuration has the property that the
variance in the number of points in an observation window $\Omega$
grows more slowly than the volume of that window \cite{ref13}.  In
the case of a spherical observation window of radius $R$, this definition
implies that the local number variance $\sigma^2(R)$ grows more
slowly than $R^d$ in $d$ dimensions, The expression for the local number
variance of a statistically homogeneous point configuration in a
spherical observation window is given exactly by
\begin{equation}\label{numvar}
\sigma^2(R) = \rho v(R) \left[1+\rho \int_{\mathbb{R}^d}
h(\mathbf{r}) \alpha(\mathbf{r}; R) d\mathbf{r}\right],
\end{equation}
where $v(R)$ is the volume of a spherical window of radius $R$ and
$\alpha(r; R)$ is the \emph{scaled intersection volume}, i.e.,
the intersection volume of two spheres of
radius $R$ separated by a distance $r$ divided by the volume of
a sphere $v(R)$. We remark that the average number of points in an
observation window is $\langle N(R)\rangle = \rho v(R)$ for any
statistically homogeneous point configuration.

It has been shown that the number variance \eqref{numvar}, under
certain conditions, admits the following asymptotic scaling
\cite{ref13}:
\begin{equation}\label{numasymp}
\sigma^2(R) = 2^d \phi\left\{A\left(\frac{R}{D}\right)^d +
B\left(\frac{R}{D}\right)^{d-1} + {\cal
O}\left[\left(\frac{R}{D}\right)^{d-3}\right]\right\},
\end{equation}
where
\begin{align}
A &= 1+\rho\int_{\mathbb{R}^d} h(\mathbf{r}) d\mathbf{r} =
\lim_{\lVert\mathbf{k}\rVert\rightarrow 0}
S(\mathbf{k})\label{An},
\end{align}
and $D$ is a characteristic microscopic length associated with the
point configuration (e.g., the average nearest-neighbor distance
between the points). Clearly, when the coefficient $A = 0$, i.e.,
$\lim_{{\bf k}\rightarrow 0}S({\bf k}) = 0$ satisfies the
requirements for hyperuniformity. The relation in \eqref{An} then
implies that hyperuniform point patterns do not possess
infinite-wavelength density fluctuations (when appropriately
scaled) and hence from (\ref{numasymp}) the number variance scales
as the surface area of the window for large $R$, i.e.,
$\sigma^2(R) \sim R^{d-1}$ in the large-$R$ limit. This result is
valid for all periodic point patterns (including perfect
crystals), quasicrystals, and disordered systems in which the pair
correlation function $g_2$ decays to unity exponentially fast
\cite{ref13}. The degree to which large-scale density fluctuations
are suppressed enables one to rank order crystals, quasicrystals
and special disordered systems \cite{ref13, chase}. Disordered
hyperuniform structures can be regarded as new states of
disordered matter in that they behave more like perfect crystals
or quasicrystals in the manner in which they suppress density
fluctuations on large length scales, and yet are also like liquids
and glasses in that they are statistically isotropic structures
with no Bragg peaks. Thus, hyperuniform disordered materials
possess a ``hidden order'' that is not apparent on short length
scales.

For disordered hyperuniform systems with a total correlation
function $h(r)$ that does not decay to zero exponentially fast,
other dependencies of the number variance on $R$ may be observed.
For example, it is known that if $S(k) \sim k$ for $k \rightarrow
0$ or, equivalently, if the total correlation function $h \sim -
r^{-(d+1)}$ for large $r$, then $\sigma^2(R) \sim (a_0\ln R + a_1)
R^{d-1}$. More generally, for any reciprocal power law,
\begin{equation}
S(k) \sim k^{\alpha} \quad (k \rightarrow 0)
\end{equation}
or, equivalently,
\begin{equation}
h(r)\sim-\frac{1}{r^{d+\alpha}} \quad (r \rightarrow +\infty),
\end{equation}
one can observe a number of different kinds of dependencies of the
asymptotic number variance $\sigma^2$ on the window radius $R$ for
$R \rightarrow \infty$ \cite{ref13, chase, chase2}:
\begin{equation}
\label{eq_S0} \sigma^2(R) \sim \left \{
\begin{array}{c@{\hspace{0.3cm}}c@{\hspace{0.3cm}}c}
R^{d-1}\ln R, & \alpha = 1, &
\\ R^{d-\alpha}, & \alpha<1, &
\\ R^{d-1}, & \alpha>1, & \end{array} \right .
\end{equation}
Note that in all cases, the number variance of a hyperuniform
point pattern grows more slowly than $R^d$.

\subsection{Order metrics}

The local bond-orientational-order metric $q_6$ is defined as
\cite{ref25}
\begin{equation}
q_6 = \left |{\frac{1}{N_b}\sum_j\sum_k \exp(6i
\theta_{jk})}\right |,
\end{equation}
where $j$ runs over all cells in the system, $k$ runs over all
neighbors of cell $j$, $\theta_{jk}$ is the angle between some
fixed reference axis in the system and the bond connecting the
centers of cells $j$ and $k$, and $N_b$ is the total number of
such bonds in the system. This quantity indicates the degree of
orientational order in the local arrangement of the immediate
neighbors of a cell and it is maximized (i.e., $q_6=1$) for the
perfect hexagonal arrangement.

To characterize translational order of a configuration, we use the
following translation order metric $T$ introduced in Ref.
\cite{order_T} and further applied in Ref. \cite{ref26},
\begin{equation}
T = \frac{1}{\eta_c}\int_0^{\eta_c}|g_2(r)-1|dr =
\frac{1}{\eta_c}\int_0^{\eta_c}|h(r)|dr
\end{equation}
where $g_2(r)$ is the pair correlation function, $h(r) = g_2(r)-1$
is the total correlation function and $\eta_c$ is a numerical
cutoff determined by the linear size of the system. The
translational order metric measures the deviation of the spatial
arrangement of cell centers in a pattern from that of a totally
disordered system (i.e., a Poisson distribution of points). The
greater the deviation from zero, the more ordered is the point
configuration.

%\subsection{Two-scale packing model}
%\subsection{Experimental data}
% Results and Discussion can be combined.
\section{Structural properties of experimentally obtained photoreceptor
patterns}

\begin{figure*}
\begin{center}
\includegraphics[width=10.5cm,keepaspectratio]{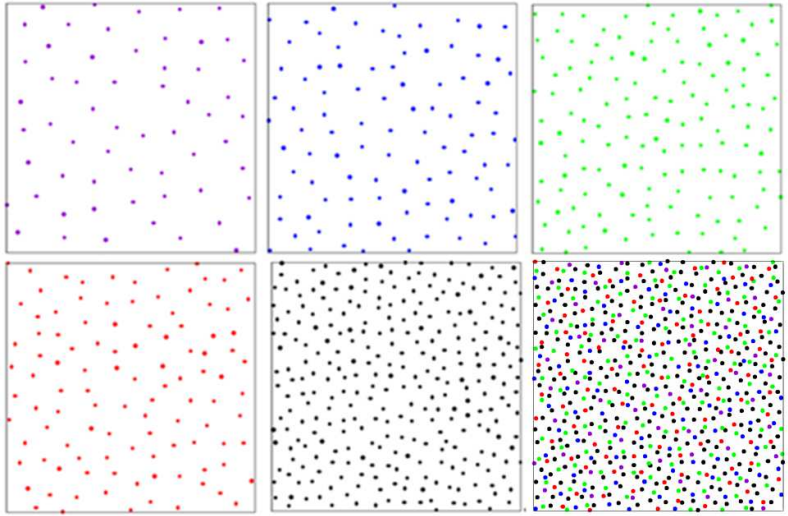}\\
\end{center}
\caption{Experimentally obtained configurations representing the
spatial arrangements of chicken cone photoreceptors. Upper panels:
The configurations shown from left to right respectively
correspond to violet, blue, green species. Lower panels: The
configurations shown from left to right respectively correspond to
red, double species and the overall pattern.}
\label{fig_cellpacking}
\end{figure*}

The chicken retina contains five different cone  cell types of
different sizes: violet, blue, green, red and double. Each cell
type of this {\it multicomponent} system is maximally sensitive to
visible light of a different wavelength. The spatial coordinates
of each cell can be determined by the presence of a colored oil
droplet in the cell's inner segment (Fig. \ref{fig1_cone}). Since
the oil droplets used to identify the locations of individual
photoreceptors are not always in exactly the same plane
\cite{ref12}, pairs of real photoreceptors sometimes appear to be
closer to one another than they are in actuality and in the
simulations. In addition, the original slightly curved retina
epithelium was flattened for imaging purposes \cite{ref12}. These
effects introduce small errors in the intercell small-distance
behavior but do not affect the overall statistics, especially on
large length scales. The spatial coordinate datasets of post-hatch
day 15 chicken (Gallus gallus) cone photoreceptors were obtained
from a published study \cite{ref12}. Each dataset contains
approximately 4430 photoreceptors, and the average numbers of
violet, blue, green, red and double species are respectively 350,
590, 880, 670 and 1840. To clearly illustrate the photoreceptor
patterns of different species, only a portion of the entire system
is shown in Fig. \ref{fig_cellpacking}. We compute a variety of
the associated statistical structural descriptors and order
metrics to quantify the degree of spatial regularity (or disorder)
of the cell arrangements.

\subsection{Disordered Hyperuniformity}

%{c@{\hspace{0.1cm}}c}
\begin{figure*}
\begin{center}
\includegraphics[height=10cm,keepaspectratio]{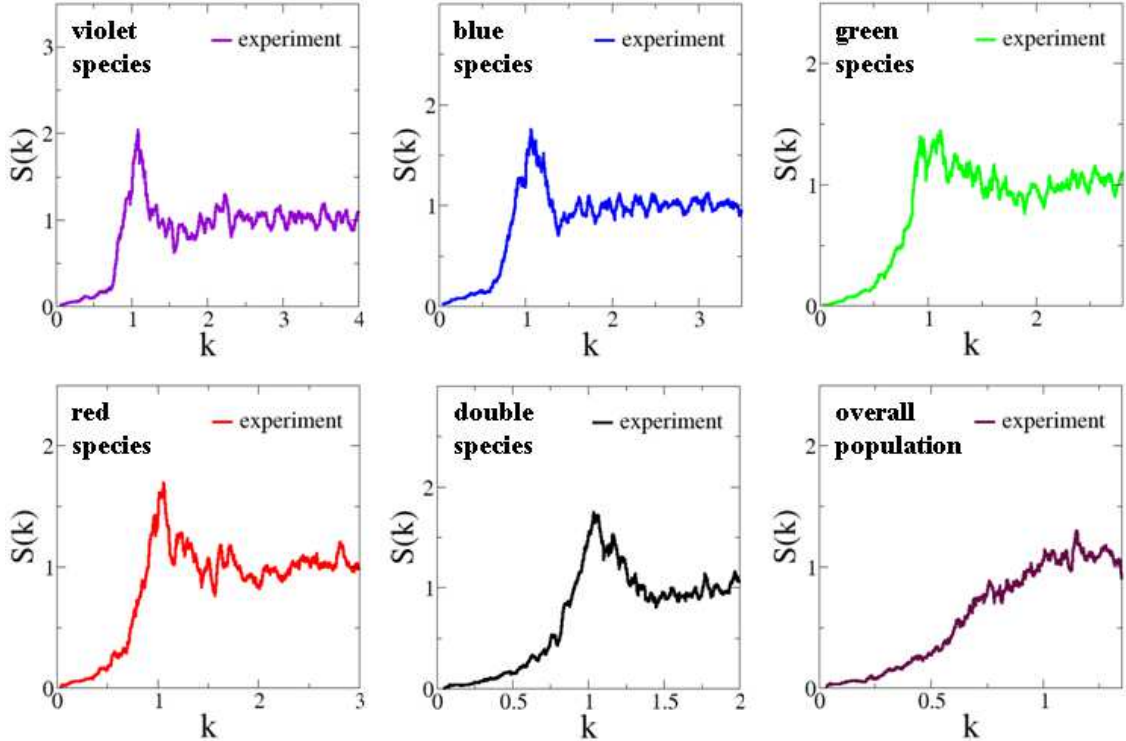} \\
\end{center}
\caption{Structure factors $S(k)$ of the experimentally obtained
point configurations representing the spatial arrangements of
chicken cone photoreceptors. The experimental data were obtained
by averaging 14 independent patterns. The estimated values of
$S(k=0)$ by extrapolation for violet, blue, green, red, double and
the overall population in the actual pattern are respectively
given by $2.11\times 10^{-3}$, $6.10\times10^{-4}$,
$1.06\times10^{-3}$, $5.72\times10^{-4}$, $1.38\times10^{-4}$,
$1.13\times10^{-3}$.} \label{fig_Sk}
\end{figure*}

As discussed in Sec. IIB, a point pattern is hyperuniform if the
number variance $\sigma^2(R)$ within a spherical sampling window
of radius $R$ (in $d$ dimensions) grows more slowly than the
window volume for large $R$, i.e., more slowly than $R^d$
\cite{ref13}. The property of hyperuniformity can also be
ascertained from the small wavenumber behavior of the structure
factor, i.e., $S(k=0)=0$ of the pattern \cite{ref13}, which
encodes information about large-scale spatial correlations (see
Sec. IIB for details). We find that $S(k)$ for the cell
configurations associated with both the total population and the
individual photoreceptor species are hyperuniform and each of
these structure factors vanishes linearly with k as k tends to
zero, i.e., $S(k) \sim k$ ($k \rightarrow 0$) (see Fig.
\ref{fig_Sk}). As discussed in Sec. IIB [cf. Eq. (\ref{eq_S0})],
such a linear behavior indicates a power-law decay for large $r$
values in the pair correlation function (i.e., $g_2(r)-1 \sim
-1/r^{3}$) instead of an exponential decay and therefore
quasi-long-range correlations in the system. We will elaborate on
this point in the ensuing discussion.

\begin{figure*}
\begin{center}
\includegraphics[height=8.5cm,keepaspectratio]{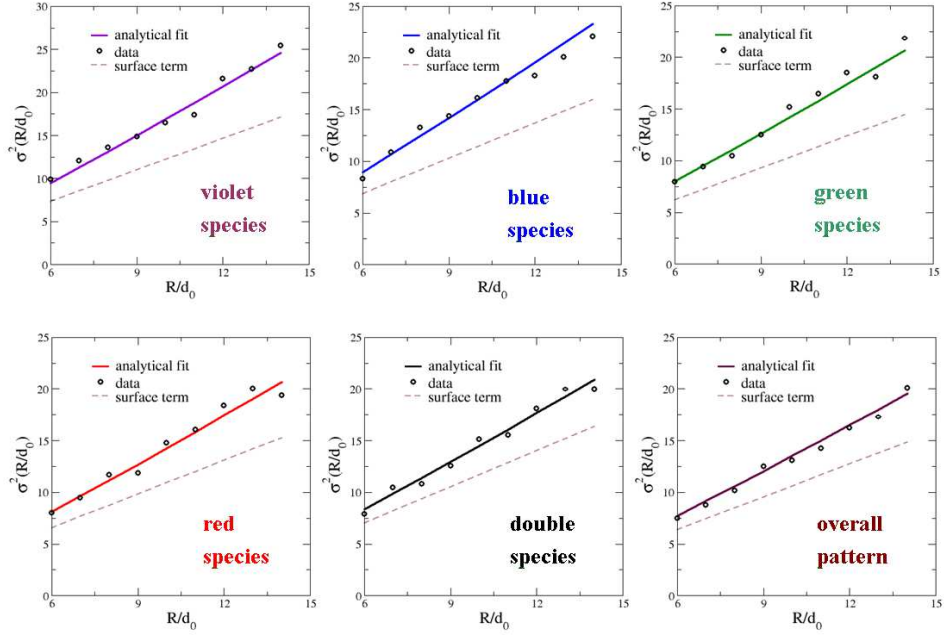} \\
\end{center}
\caption{The number variance $\sigma^2(R)$ associated with the
photoreceptor patterns in chicken retina as well as the associated
fitting function of the form $\sigma^2(R) = A R^2 + B R\ln(R) + C
R$. We found that the values of the parameter $A$ are several
orders of magnitude smaller than the other two parameters,
indicating that the associated patterns are effectively
hyperuniform. Also shown in each plot is the ``surface term'' $C
R$ for purposes of comparison. The window radius $R$ is normalized
with respect to the mean nearest neighbor distance $d_0$ of the
corresponding point configurations.} \label{fig_sigma}
\end{figure*}

We have directly computed the number variance $\sigma^2(R)$ for
the individual and overall patterns and verified that they are
also consistent with hyperuniformity, i.e., the ``volume term'' in
$\sigma^2(R)$ is several orders of magnitude smaller than the
other terms [c.f. Eq.~\eqref{numasymp}]. Specifically, for each
$R$ value, 2500 windows are randomly placed in the system without
overlapping the system boundary. The finite system size $L$
imposes an upper limit on the largest window size, which is chosen
to be $R_{max} = L/2$ here. Figure \ref{fig_sigma} shows the
experimental data as well as the associated fitting functions of
the form
\begin{equation}
\sigma^2(R) = A R^2 + B R\ln(R) + C R,
\end{equation}
where $A = S(k=0)$ and $B, C>0$. Note that in the plots, the
window size $R$ is normalized by the corresponding
nearest-neighbor distance $d_0$ for each species. Also shown in
each plot is the corresponding ``surface term'' $C R$ for purposes
of comparison. The numerical values of the fitting parameters for
both the overall pattern and the individual species are given in
Table \ref{tab_1}. It can be clearly seen that the values of the
parameter $A$ are several orders of magnitude smaller than the
other two parameters, indicating that the associated patterns are
effectively hyperuniform. These values are also consistent with
the numerical values of $S(k=0)$ obtained by directly fitting
$S(k)$ for small $k$ values \cite{footnote0}.

\begin{table*}[h]
\caption{The numerical values of the fitting parameters for both
the overall pattern and the individual species.}
\begin{tabular}{c|@{\hspace{0.35cm}}c@{\hspace{0.35cm}}c@{\hspace{0.35cm}}c@{\hspace{0.35cm}}c@{\hspace{0.35cm}}c@{\hspace{0.35cm}}c}
\hline
&  Violet & Blue &   Green  & Red & Double & Overall \\
\hline $A$ &  $2.53\times 10^{-4}$ &   $9.24\times 10^{-4}$ &
$1.07\times 10^{-3}$ & $1.77\times 10^{-3}$ & $4.46\times 10^{-3}$
& $1.93\times 10^{-3}$ \\
$B$ &  0.203 &  0.198 &  0.169 &  0.146 &  0.122 &  0.127\\
$C$ & 1.22 & 1.14 &  1.03 &  1.09  &  1.17  &   1.06 \\
\hline
\end{tabular}
\label{tab_1}
\end{table*}

The fact that the photoreceptor patterns display both overall
hyperuniformity and homotypic hyperuniformity implies that if any
subpopulation of the individual species is removed from the
overall population, the remaining pattern is still hyperuniform.
We term such patterns {\it multi-hyperuniform} because distinct
multiple subsets of the overall point pattern are themselves
hyperuniform. These are highly unusual and unique structural
attributes. Until now, the property of \textit{overall}
hyperuniformity was identified only in a special subset of
disordered physical systems \cite{ref15, ref16, ref18, berthier,
weeks, aleksPRL, jiaoPRE, helium, plasma,universe, fermion,
ref17}. The chicken photoreceptor patterns provides the first
example of a disordered hyperuniform biological system. In
addition, the photoreceptor patterns possess quasi-long-range
(QLR) correlations as indicated by the linear small-$k$ behavior
in $S(k)$. We will elaborate on these points in Sec. V.

\subsection{Pair Correlation Functions}

\begin{figure*}
\begin{center}
\includegraphics[height=10cm,keepaspectratio]{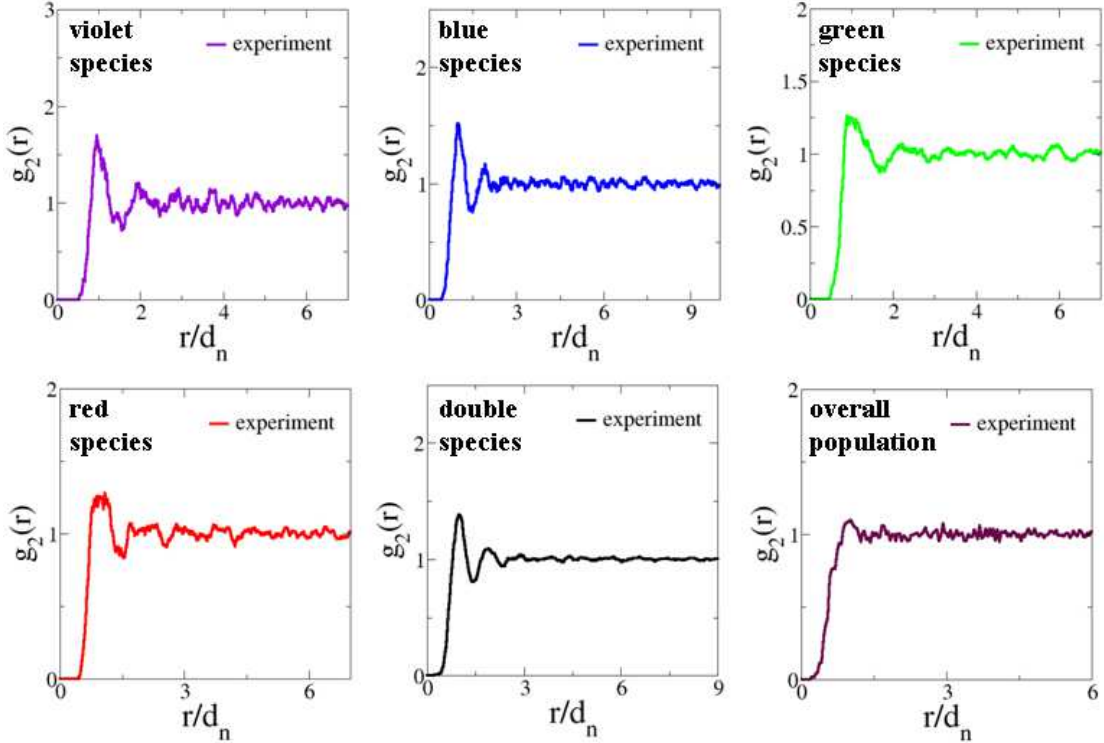} \\
\end{center}
\caption{Pair correlation functions $g_2(r)$ of the experimentally
obtained point configurations representing the spatial
arrangements of chicken cone photoreceptors. The experimental data
were obtained by averaging 14 independent patterns. The distance
is rescaled by the average nearest neighbor distance $d_n$ in the
system.} \label{fig_g2}
\end{figure*}

We find that each cell is associated with an effective exclusion
region (i.e., an area in 2D) with respect to any other cells,
regardless of the cell types. The size of these exclusion regions
roughly corresponds to the size of the cells themselves
\cite{ref12}. In addition, cells belonging to the same subtype
(i.e., like-cells) are found to be mutually separated from one
another almost as far as possible, leading to a larger effective
exclusion region associated with like-cells of each species. The
exclusion effects are quantitatively captured by the associated
pair-correlation functions (Fig. \ref{fig_g2}). The hard-core
exclusion effect is manifested in $g_2(r)$ as an interval of $r$
for which $g_2(r) = 0$ (i.e., an ``exclusion gap'') and $g_2(r)$
approaches its large-$r$ asymptotic value of unity very quickly,
indicating the absence of any long-range spatial ordering. This is
to be contrasted with ordered systems, such as crystals, whose
pair correlation functions are composed of individual Dirac delta
functions at specific $r$ values.

\subsection{Order Metrics}

%insert a table here
\begin{table*}[h]
\caption{Bond-orientational and translational order metrics, $q_6$
and $T$, respectively, of the chicken photoreceptor patterns. The
experimental data were obtained by averaging 14 independent
patterns.}
\begin{tabular}{@{\vrule height 10.5pt depth4pt  width0pt}c|c|c}
\hline
~Species~ &  ~$q_6$~ & ~$T$~ \\
% \hline
\hline
Violet & ~0.150~  & ~0.304~ \\
Blue & ~0.158~  & ~0.411~  \\
Green & ~0.130~  & ~0.278~  \\
Red & ~0.147~  & ~0.254~  \\
Double & ~0.184~  & ~0.390~ \\
All & ~0.058~  & ~0.096~  \\
\hline
\end{tabular}
\label{tab_2}
\end{table*}

A bond-orientational order metric $q_6$ \cite{ref25} and a
translational order metric $T$ \cite{ref26} were  used next to
quantify the degree of spatial regularity in the photoreceptor
patterns (see Tab. \ref{tab_2}), each of which are maximized by
the triangular lattice and minimized by a spatially uncorrelated
point pattern. Interestingly, the $q_6$ and $T$ values for the
total population are close to the corresponding values for
polydisperse hard-disk packings we obtained, implying that local
cell exclusion effect plays a primary role in determining the
overall pattern. In contrast, the higher $q_6$ and $T$ values for
individual cell species suggest that like-cells interact with one
another on a length scale larger than the size of a single cell,
which tends to increase the degree of order in the arrangements of
like-cells.

From a functional point of view, photoreceptor cells of a given
type maximize their sampling efficiency when arranged on an
ordered triangular lattice, as in the case of the compound eye of
insects \cite{ref5, ref6}. Importantly, the triangular lattice has
been shown to be the most hyperuniform pattern \cite{ref13}, i.e.,
it minimizes the large-scale density fluctuations among all 2D
patterns. However, this most hyperuniform pattern may not be
achieved if other constraints  (e.g., cell size polydispersity)
are operable. We therefore hypothesize that the disordered
hyperuniformity of avian photoreceptor patterns represents a
compromise between the tendency of the individual cell types to
maximize their spatial regularity and the countervailing effects
of packing heterotypic cell types within a single epithelium,
which inhibits the spatial regularity of the individual cell
types. In other words, the avian photoreceptors are driven to
achieve the most ``uniform'' spatial distribution subject to
heterotypic cell packing constraints.

\section{Computational Model That Yields Multi-Hyperuniform Patterns}

Our initial attempt to model the avian photoreceptor cell patterns
employed classic packing models of polydisperse hard disks that
are driven to their ``jammed states'' \cite{ref27}. However, these
models failed to generate patterns with multi-hyperuniformity. Such standard
jamming models involving interactions on a single length scale are
insufficient to represent the two competing effects leading to the
photoreceptor patterns and motivated us to develop a unique
multiscale packing model as described below.

\begin{figure*}
\begin{center}
\includegraphics[width=10.5cm,keepaspectratio]{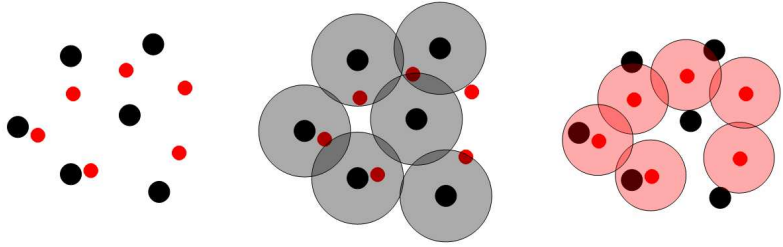}
\end{center} \caption{Illustration of the hard-core and soft-core
interactions in a two-species system containing black and red
cells. The left panel shows the exclusion regions (circular disks
with two distinct sizes) associated with the two types of cells,
which is proportional to the actual sizes of the cells. The black
cells have a larger exclusion region than the red cells. The
middle panel illustrates the soft-core repulsive interaction
(large concentric overlapping circles of the solid black disks)
between the black cells. Such a repulsive interaction will drive
the black cells to arrange themselves in a perfect triangular
lattice in the absence of other species. The right panel
illustrates the soft-core repulsive interaction (large concentric
overlapping circles of the solid red disks) between the red
cells.} \label{fig_packing}
\end{figure*}

In the experimental data representing the spatial arrangements of
chicken cone photoreceptors, each cell is represented by a point.
We refer to these points as ``cell centers'', although they may
not correspond to the actual geometrical centers of the cells.

In order to modify a simple hard-core interaction, we consider two
types of effective cell-cell interactions: isotropic short-range
hard-core repulsions between any pair of cells and isotropic
long-range soft-core repulsions between pairs of like-cells (i.e.,
cells of the same subtype). The multiscale nature of the model
results from the multiple length scales involved in these
interactions for different species, as we discuss now. The
strength of the hard-core repulsion is characterized by the radius
$R^i_h$ of a hard-disk exclusion region associated with a cell
type $i$. This interaction imposes a nonoverlap constraint such
that the distance between the cells $i$ and $j$ cannot be smaller
than $(R^i_h + R^j_h)$, which mimics the physical cell packing
constraint. In this regard, $R^i_h$ will also be referred to as
the radius of a cell $i$ in the ensuing discussions. The relative
magnitudes of $R^i_h$ are estimated from an electron micrograph
showing photoreceptor cell packing at the level of the inner
segment (see discussion below) \cite{ref11}. The characteristic
radius $R_s$ of the soft-core repulsion is associated with the
mean nearest-neighbor distance of the cells of the same type.
Specifically, the pair potential between two like-cells is given
by
\begin{equation}
\label{potential} E(r) =
\left\{{\begin{array}{c@{\hspace{0.8cm}}c}
\displaystyle{\frac{\alpha}{\beta+1}}(2R_s-r)^{\beta+1} & r\le
2R_s,
\\\\ 0 & r>2R_s, \end{array}} \right .
\end{equation}
where the parameters $\alpha>0$ and $\beta>0$ set the scale of the
interaction energy \cite{footnote2}. In our simulations, we
require that value of $R_s$ be uniquely determined by the
associated cell number density $\rho$, i.e., $R_s =
\frac{1}{2}\sqrt{2/(\sqrt(3)\rho)}$. This implies that a system
composed of cells of the same type (i.e., a single-component
system) interacting via a pair potential given by Eq.
\eqref{potential} at number density $\rho$ (i.e., the number of
cells per unit area) possesses the triangular-lattice ground
state, i.e., an arrangement associated with a minimal total energy
(sum of the total interaction energy between any pairs of
like-cells). In other words, when the total energy in a
single-component system is reduced to its minimal value (e.g.,
zero), sufficiently slowly from an arbitrary initial
configuration, the cells will reorganize themselves into a
triangular-lattice arrangement.

When the system contains multiple species, the hard-core and
soft-core interactions represent two competing effects in
determining the packing arrangement of the cells; see Fig.
\ref{fig_packing}. Specifically, the polydisperse hard-disk
exclusion regions induce geometrical frustration in the packing,
i.e., in this five-component system, it is not possible for the
subset of disks with the same size, surrounded by disks with
different sizes to be arranged on a perfect triangular lattice. On
the other hand, the long-range soft interaction between like
species tends to drive the cells of the same type to arrange
themselves on a perfect triangular lattice. Note that although the
relative magnitudes of $R^i_h$ for different species (i.e., the
ratio between any two $R^i_h$) are fixed, the actual values of
$R^i_h$ are variable and used as a tuning parameter in our model.
As stated above, the ratios between $R^i_h$ are estimated from a
previously published study \cite{ref11}. Specifically, the
relative sizes of the violet, blue, green, red and double species
are 1.00, 1.19, 1.13, 1.06 and 1.50, respectively. Given the
number of cells of each species, the values of $R^i_h$ can be
uniquely determined from the packing fraction $\phi$ of the cells
(i.e., the fraction of space covered by the cells) and vice versa,
\begin{equation}
\label{eq_phi}
\phi = \displaystyle{\frac{1}{A}\sum_i N_i \pi (R^i_h)^2},
\end{equation}
where $N_i$ is the number of cells of species $i$ and $A$ is the
area of the system.

Our Monte Carlo algorithm, which involves iterating ``growth'' and ``relaxation'' steps,
works as follows:

\begin{itemize}

\item{(1) Initialization. In the beginning of the simulation, cell centers of each species are generated in a
simulation box using the random-sequential-addition (RSA) process
\cite{ref27}. Specifically, for each species $i$, $N_i$ cell
centers are randomly generated such that these cell centers are
mutually separated by a minimal distance $\mu R_s$ $(0<\mu<1)$. In addition,
the newly added cell cannot overlap any existing cells in the box
(determined by the hard-core radius $R_s$), regardless of cell types. The initial covering fraction $\phi_I$
associated with the hard-core exclusion regions is determined by
$R^i_h$ via Eq. (\ref{eq_phi}), and is about $80\%$ of the RSA saturation density \cite{ref27}.}

%%%%%%%%%%%%%%%%%%%%%%%%%%%%%%%%%%%%%%%%%%%%%%%%%%%%%%%%%%%%%%%%%%%

\item{(2) Growth step. At each stage $n$, the cells are allowed to randomly move a prescribed
maximal distance ($\sim 0.25 R^i_h$) and direction such that no
pairs of cells overlap. After a certain number ($\approx$1,000) of
such random movements for each cell, the radius $R^i_h$ of each
cell is increased by a small amount such that the size ratios of
the cells remain the same. This leads to an increase of the
packing fraction $\phi_n$ at this stage by an amount of about $1\% - 3\%$.
Note that in this ``growth'' step, the long-range
soft interactions between the like-cells are turned off.}

\item{(3) Relaxation step. At the end of the ``growth'' step,
the soft interactions are then turned on, and the cells are
allowed to relax from their current positions to reduce the total
system energy subject to the nonoverlap condition. The steepest
decent method is used to drive the system to the closest local
energy minimum (i.e., the inherent structure \cite{ref27}) associated with the starting configuration. This is
referred to as the ``relaxation'' process.}

%%%%%%%%%%%%%%%%%%%%%%%%%%%%%%%%%%%%%%%%%%%%%%%%%%%%%%%%%%%%%%%%%%%

\item{(4) Statistics. After the relaxation process, structural statistics of the
resulting configuration of cell centers are obtained and compared
to the corresponding experimental data. To ensure that the simulations
match the data for the pair statistics to the best extent possible,
we introduce a deviation metric $\Delta$. Specifically, $\Delta$ is the
normalized sums of the squared differences between the simulated and experimental $S(k)$
and $g_2(r)$ associated with the simulated and actual patterns, i.e.,
\begin{equation}
\label{eq_delta}
\Delta = \frac{1}{n_S}\sum_i^{n_S}\sum_r[g^{(i)}_2(r) - \bar{g}^{(i)}_2(r)]^2
 +\frac{1}{n_S}\sum_i^{n_S}\sum_k[S^{(i)}(k) - \bar{S}^{(i)}(k)]^2,
\end{equation}
where $n_S = 6$ is the total number of species including both the 5 individual species
and the overall pattern, $g_2^{(i)}(r)$ and $S^{(i)}(k)$ are the simulated functions associated with
species $i$, and $\bar{g}_2^{(i)}(r)$ and $\bar{S}^{(i)}(k)$ are the corresponding
experimentally measured functions.}

\item{(5) The growth and relaxation steps described in the bullet items (2) and (3),
respectively, are repeated until $\phi_n$ reaches a prescribed
value $\phi_F$. Specifically, the configuration obtained by relaxation at stage $n$ is used
as the starting point for the growth step at stage $n+1$. The best simulated pattern (i.e., that with the smallest
deviation metric $\Delta_{min}$) and the associated $\phi^*$ value are
then identified.}

\end{itemize}

At a given packing fraction $\phi$ (or equivalently a set of
$R^i_h$), the polydispersity of the exclusion regions associated
with different species and the resulting nonoverlap constraints
frustrate the spatial order in the system. For example, the
long-range soft interaction drives a single-species system to the
triangular-lattice arrangement in the absence of other species. On
the other hand, for any $\phi >0$, it is impossible for cells of a
particular species, surrounded by cells of other species to sit on
a perfect triangular lattice \cite{ref12}. Therefore, the
disordered point configurations obtained by minimizing the energy
associated with the soft repulsive interactions subject to the
hard-core packing constraints are the local energy minima (i.e., inherent structures)
of the system. The extent to which the structure deviates from that of a
perfect triangular lattice (i.e., global energy minimum) is
determined by the parameter $\phi$ (or, equivalently, $R^i_h$).
Therefore, by tuning this parameter in our algorithm, one can, in
principle, generate a continuous spectrum of configurations of
cell centers with varying degrees of spatial order (see Appendix).
Note that in the limit $R^i_h \rightarrow 0$, triangular-lattice
arrangements for individual species are accessible again and the
resulting configuration is a superposition of five
triangular-lattice arrangements of the cell centers.

We note that the order of the aforementioned growth and relaxation
steps can be interchanged without affecting the final
configuration. In addition, instead of starting from a disordered
RSA arrangement of cell centers as described above, we have also
used ordered initial configurations (i.e., superposition of
triangular-lattice arrangements), leading to the same
configuration at a given number density $\rho$. However, the
initial packing density $\phi_I$ associated with ordered initial
configurations is very low and thus, it is computationally
inefficient to start from such initial configurations. By tuning
the ``strength'' of the hard-core interactions via the packing
fraction associated with the exclusion regions, our multiscale
packing model enables us to produce disordered point configuration
with various degrees of hyperuniformity, examples of which are
provided in the Appendix for a three-component system for
illustrative purposes.

\subsection{Modeling Avian Photoreceptor System via Multiscale
Particle Packing}

\begin{figure*}
\begin{center}
\includegraphics[width=11.5cm,keepaspectratio]{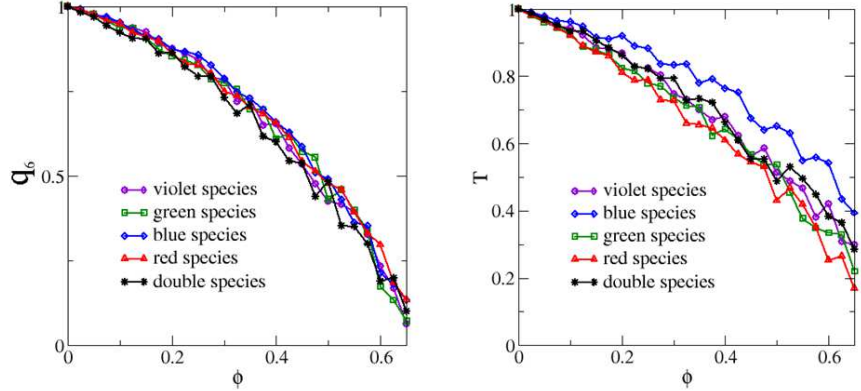} \\
\end{center}
\caption{Left panel: The bond-orientational order metric $q_6$ of
the individual species as a function of the packing fraction
$\phi$ associated with the exclusion regions. Right panel: The
translational order metric $T$ of the individual species as a
function of the packing fraction $\phi$ associated with the
exclusion regions.} \label{fig_order}
\end{figure*}

By using the multiscale packing model, we were able to accurately
reproduce the unique features of the native avian photoreceptors.
We modeled the aforementioned two competing effects as two types
of effective interactions between the cells: a long-range
soft-core repulsion between the cells of the same type (that would
lead to an ordered triangular-lattice arrangement in the absence
of packing constraints) and a short-range hard-core repulsion
(with polydisperse exclusion regions associated with different
cell species) between any pair of cells that frustrates spatial
ordering in the system. Given the sizes of the hard-core exclusion
regions associated with each cell species (or equivalently the
packing fraction $\phi$ of the exclusion regions), the system is
allowed to relax to a state that is a local energy minimum for the
long-range soft-core repulsive interactions between like-species.
Such long-range interactions would drive each of the five cell
species in the multicomponent system to the associated
triangular-lattice arrangement (global energy minimum) in the
absence of the hard-core repulsions. As we increase the strength
of the hard-core repulsions by increasing $\phi$, the degree of
order in the system, which is quantified by the order metrics
$q_6$ and $T$, decreases (see Fig. \ref{fig_order}). It is
important to emphasize that these disordered hyperuniform avian
photoreceptor patterns are {\it not} simple random perturbations
of a triangular-lattice pattern. Statistically equivalent
disordered hyperuniform patterns have also been obtained from
disordered initial configurations (e.g., RSA packings). Thus, the
unique structural features in these patterns are not attributed to
particular initial configurations but rather arise from the two
competing effects, which are well captured by our multiscale
packing model.

\begin{figure*}
\begin{center}
\includegraphics[width=10.5cm,keepaspectratio]{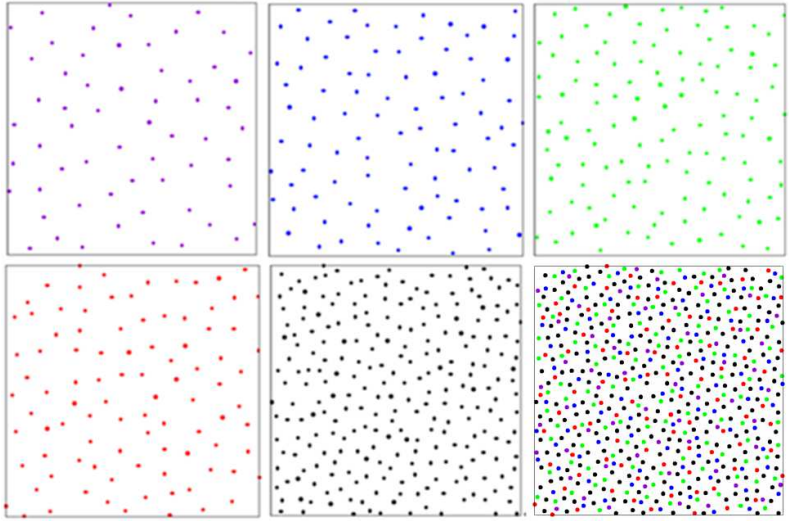}\\
\end{center}
\caption{Simulated point configurations representing the spatial
arrangements of chicken cone photoreceptors. Upper panels: The
configurations shown from left to right respectively correspond to
violet, blue, green species. Lower panels: The configurations
shown from left to right respectively correspond to red, double
species and the overall pattern. The simulated patterns for
individual photoreceptor species are virtually indistinguishable
from the actual patterns obtained from experimental measurements.}
\label{fig_simupacking}
\end{figure*}

%{c@{\hspace{0.1cm}}c}
\begin{figure*}
\begin{center}
\includegraphics[height=10cm,keepaspectratio]{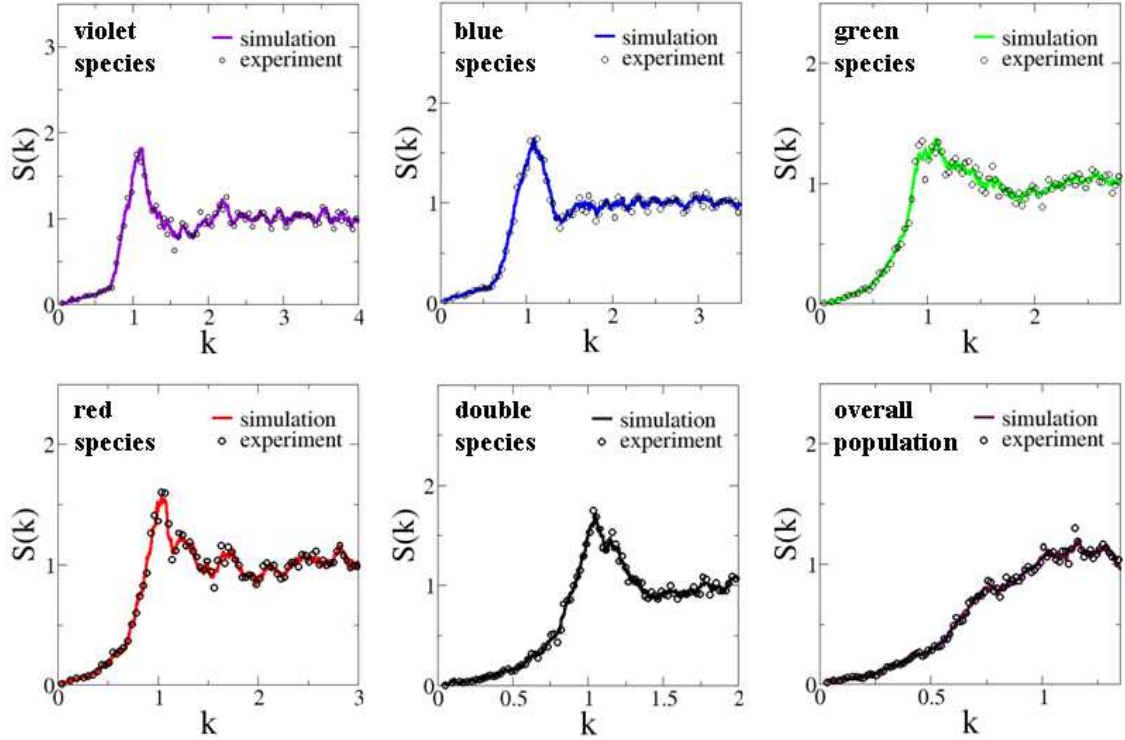} \\
\end{center}
\caption{Comparison of the structure factors $S(k)$ of the
experimentally obtained and simulated point configurations
representing the spatial arrangements of chicken cone
photoreceptors. The simulation data were obtained by averaging 50
independent configurations.} \label{fig_simuSk}
\end{figure*}

\begin{figure*}
\begin{center}
\includegraphics[height=10cm,keepaspectratio]{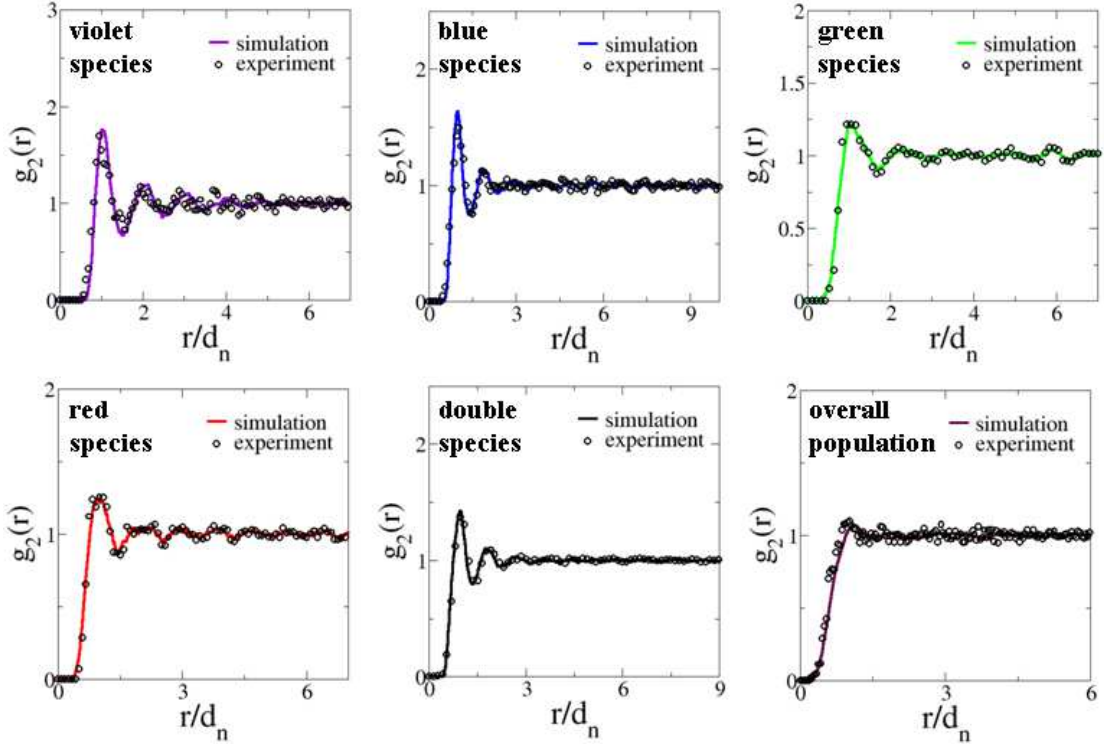} \\
\end{center}
\caption{Comparison of the pair correlation functions $g_2(r)$ of
the experimentally obtained and simulated point configurations
representing the spatial arrangements of chicken cone
photoreceptors. The simulation data were obtained by averaging 50
independent configurations. The distance is rescaled by the
average nearest neighbor distance $d_n$ in the system.}
\label{fig_simug2}
\end{figure*}

\begin{table*}[h]
\caption{Comparison of the bond-orientational and translational
order metrics, $q_6$ and $T$, of the experimental and simulated
point configurations. The simulation data were obtained by
averaging 50 independent configurations.}
\begin{tabular}{@{\vrule height 10.5pt depth4pt  width0pt}c|c|c|c|c }
\hline
&  \multicolumn{2}{|c|}{$q_6$} & \multicolumn{2}{|c}{$T$} \\
% \hline
~Species~& ~Exp.~ & ~Sim.~ & ~Exp.~ & ~Sim.~ \\
\hline
Violet & 0.150 &  0.148 & 0.304 &  0.327 \\
Blue & 0.158 &  0.164 & 0.411  & 0.395 \\
Green & 0.130 &  0.134 & 0.278 &  0.266 \\
Red & 0.147 &  0.149 & 0.254  & 0.263 \\
Double & 0.184 &  0.189 & 0.390 &  0.363\\
All & 0.058 &  0.063 & 0.096  & 0.108 \\
\hline
\end{tabular}
\label{tab_simu}
\end{table*}

%We find that the simulated configurations are hyperuniform over a
%wide range of packing fractions of the exclusion regions, i.e.,
%$\phi \in [0, 0.63]$.

The simulation box contains 2600 cell centers and the
numbers of violet, blue, green, red and double species are
respectively 210, 355, 530, 405, and 1100. The
relative sizes of the violet, blue, green, red and double species
are 1.00, 1.19, 1.13, 1.06 and 1.50, respectively.
The initial packing fraction associated with the hard cores
is $\phi_I = 0.45$ and the simulation stops at $\phi_F = 0.7$.
At $\phi \approx 0.58$, the resulting
configurations (see Fig. \ref{fig_simupacking}) are virtually
indistinguishable from the actual photoreceptor patterns, as
quantified using a variety of descriptors. Specifically, the
associated structure factors (see Fig. \ref{fig_simuSk}) and pair
correlation functions (see Fig. \ref{fig_simug2}) match the
experimental data very well, as quantified by the minimum
deviation metric value of $\Delta_{min} \approx 0.4$ [c.f. Eq.(\ref{eq_delta})].
We note that the major contributions to $\Delta_{min}$ are the large
fluctuations in the experimental data due to a limited number of samples.
(The initial value of $\Delta$ is roughly $3.16$.)
The order metrics $q_6$ and $T$ of the simulated pattern also match
those of the experimental data very well (see Tab. \ref{tab_simu}).
This is a stringent test for the simulations to pass. The success
of the simulations strongly suggests that the disordered
hyperuniform photoreceptor patterns indeed arise from the
competition between cell packing constraints and the tendency to
maximize the degree of regularity for efficient light sampling,
suggesting that the individual photoreceptor types are as uniform
as they can be, given the packing constraints within the
photoreceptor epithelium.

\section{Conclusions and Discussion}

By analyzing the chicken cone photoreceptor patterns using a
variety of sensitive microstructural descriptors arising in
statistical mechanics and particle-packing theory, we found that
these disordered patterns display both overall and homotypic
hyperuniformity, i.e., the system is multi-hyperuniform. This
singular property implies that if any subset of the individual
species is removed from the overall population, the remaining
pattern is still hyperuniform. Importantly, it is highly
nontrivial to devise an algorithm that would remove a large
fraction of the points from a disordered hyperuniform system while
leaving the remaining point pattern hyperuniform, and yet Nature
has found such a design.

Until now, the property of \textit{overall} hyperuniformity was
identified only in a special subset of disordered physical
systems, including ground-state liquid helium \cite{helium},
one-component plasmas \cite{plasma}, Harrison-Zeldovich power
spectrum of the density fluctuations of the early Universe
\cite{universe}, fermionic ground states \cite{fermion}, classical
disordered ground states \cite{ref17}, and maximally random jammed
packings of equal-sized hard particles \cite{aleksPRL, jiaoPRE}.
All of these examples involve single-component systems. More
recently, disordered multicomponent physical systems such as
maximally random jammed (MRJ) hard-particle packings \cite{ref16,
ref18, ref15} have been identified that possess an appropriately
generalized hyperuniformity property ascertained from the local
volume fraction fluctuations. However, the multicomponent
photoreceptor avian system pattern, which represents the first
example of a disordered hyperuniform system in a living organism,
is singularly different from any of these hyperuniform physical
systems in that in the pattern each species and the total
population are hyperuniform, i.e., the avian patterns are
multi-hyperuniform. Although it is not very difficult to construct
a overall hyperuniform system by superposing subsystems that are
individually hyperuniform, the reverse process (i.e., decomposing
a hyperuniform system into individually hyperuniform subsets) is
highly nontrivial. It will be of interest to identify other
disordered hyperuniform biological systems. It is likely that some
other epithelial tissues and phyllotactic systems \cite{ref27}
possess such attributes. Interestingly, it has been shown that the
large-scale  number-density fluctuations associated with the
malignant cells in brain tumors are significantly suppressed,
although the cell patterns in such brain tumors are not
hyperuniform \cite{plos}.

In addition, the photoreceptor patterns possess quasi-long-range
(QLR) correlations as indicated by the linear small-$k$ behavior
in $S(k)$. Such QLR correlations are also observed in the
ground-state liquid helium \cite{helium}, the density fluctuations
of the early Universe \cite{universe}, fermionic ground states
\cite{fermion} and MRJ packings of hard particles \cite{ref16,
ref18, ref15}. In the MRJ particle packings, it is believe that
the QLR correlations arise from the competition between the
requirement of jamming and maximal disorder in the system
\cite{ref16, ref18, ref15}. As we showed employing the unique
multiscale packing model, the multicomponent avian system that is
both homotypic and overall hyperuniform, i.e., multi-hyperuniform,
can result from two competing interactions between the
photoreceptors.

\begin{figure*}
\begin{center}
\includegraphics[width=10.5cm,keepaspectratio]{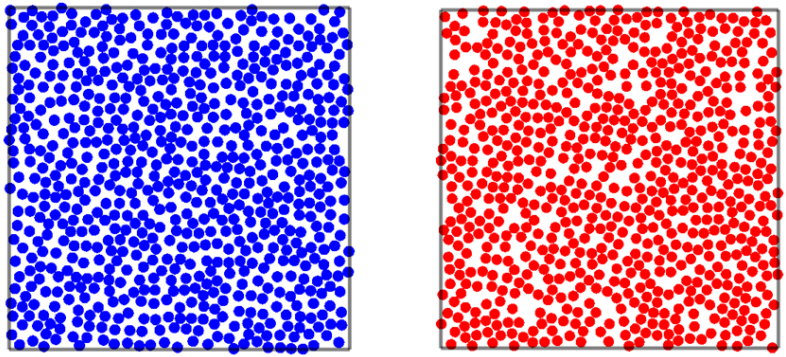}\\
\end{center}
\caption{Left panel: A random-sequential-addition (RSA) packing of
hard, identical circular disks in two-dimensions with a packing
fraction $\phi = 0.54$, which is close to the saturation state.
Right panel: An equilibrium system of hard, identical disks at
$\phi = 0.54$. The fact that neither of these systems is
hyperuniform, as discussed in the text, indicates that hard-core
exclusion effects alone are not sufficient to induce
hyperuniformity.} \label{fig2_rsa}
\end{figure*}

It is noteworthy that while hard-core exclusion and high density
in a disordered particle packing are necessary conditions to
achieve a hyperuniform state, these are not sufficient conditions.
Figure \ref{fig2_rsa} shows a nonequilibrium
random-sequential-addition (RSA) packing of hard circular disks in
two-dimensions with a packing fraction $\phi = 0.54$ (left panel),
which is generated by randomly and sequentially placing hard disks
in a domain without overlapping existing disks, until there is no
room for additional disks \cite{rsa}. The right panel of Fig.
\ref{fig2_rsa} shows an equilibrium system of hard disks at $\phi
= 0.54$ (right panel). The structure factor values at $k=0$ for
the RSA and equilibrium systems are respectively given by $S(0)=
0.059$ \cite{rsa} and $S(0) = 0.063$ \cite{salbook, newref43,
footnote}. Although hard-core exclusion play a central role in
these two distinct high-density packings, neither packing is
hyperuniform, as indicated by the relatively large positive values
of the corresponding $S(0)$.

%we first considered classic packing models of polydisperse hard
%disks, which failed to generate patterns with overall and
%homotypic hyperuniformity, indicating that standard single-scale
%packing models cannot capture the salient features of the
%photoreceptor patterns.} This also motivated us to

To understand the origin of the unique spatial features of the
avian photoreceptor patterns, we have devised a unique multiscale
cell packing model that suggests that photoreceptor types interact
with both short- and long-ranged repulsive forces and that the
resultant competition between the types gives rise to the singular
cell patterns. The fact that a disordered hyperuniform pattern
corresponds to a local optimum associated with the multiscale
packing problem indicates that such a pattern may represent the
most uniform sampling arrangement attainable in the avian system,
instead of the theoretical optimal solution of a regular hexagonal
array. Specifically, our studies show how fundamental physical
constraints can change the course of a biological optimization
process. Although it is clear that physical cell packing
constraints are the likely cause of the short-range hard-core
repulsion, the origin of the effective longer-range soft-core
repulsion is less obvious. We hypothesize that repulsive forces of
this type occur during retinal development and may be secondary to
cell-cell interactions during photoreceptor neurogenesis. However,
a comprehensive test of this hypothesis is beyond the scope of
this investigation, and therefore its resolution represents a
fascinating avenue for future research.

Recent studies have shown that disordered hyperuniform materials
can be created that possess unique optical properties, such as
being ``stealthy'' (i.e., transparent to incident radiation at
certain wavelengths) \cite{ref17}. Moreover, such disordered
hyperuniform point patterns have been employed to design isotropic
disordered network materials that possess complete photonic band
gaps (blocking all directions and polarizations of light)
comparable in size to those in photonic crystals \cite{ref28,
man13}. While the physics of these systems are not directly
related to the avian photoreceptor patterns, such investigations
and our present findings demonstrate that a class of disordered
hyperuniform materials are endowed with novel photonic properties.

Besides capturing the unusual structural features of photoreceptor
patterns, our multiscale packing model represents a unique
algorithm that allows one to generate multi-hyperuniform
multicomponent systems with varying degrees of order by tuning the
packing fraction $\phi$ of the hard-core exclusion regions (see
Appendix for additional examples). This knowledge could now be
exploited to  produce multi-hyperuniform disordered structures for
applications in condensed matter physics and materials science. For example, it would be of
interest to explore whether colloidal systems can be synthesized
to have such repulsive interactions in order  to self assemble
into the aforementioned unique disordered arrangements and to study
the resulting optical properties. It is noteworthy that it has
already been demonstrated that three-dimensional disordered hyperuniform
polymer networks can be fabricated for photonic applications using direct
laser writing \cite{polymer}.
% Additional
%applications of this approach would be the development of image
%sensors \cite{ref29} with an arbitrary number of color channels or
%an ability to scatter light (or other waves) at selective
%frequencies depending on the nearest-neighbor distances between
%the different species, which are by themselves hyperuniform.

%comment on the protocol limitations in producing such packings and possible potential protocols
%comment geometric-point of view of characterization particles is superior in the sense
% that it can incorporate interesting packing structures vs. the ensemble point of view,
% use the same 2D argument.

\begin{acknowledgments}
The authors are grateful to Paul Steinhardt for useful
discussions. Y. J. and S. T. were supported by the National Cancer
Institute under Award NO. U54CA143803 and by the Division of
Mathematical Sciences at the National Science Foundation under
Award No. DMS-1211087. J.C.C. was supported by NIH grants
(EY018826, HG006346 and HG006790) and J.C.C., M.M.-H. and H.H.
were supported by a grant from the Human Frontier Science Program.
H.H. also acknowledges the support of the German Research
Foundation (DFG) within the Cluster of Excellence, ``Center for
Advancing Electronics Dresden''. This work was partially supported
by a grant from the Simons Foundation (Grant No. 231015 to
Salvatore Torquato).
\end{acknowledgments}

\appendix

\renewcommand{\theequation}{A-\arabic{equation}} % redefine the command that creates the equation no.
\setcounter{equation}{0}  % reset counter

\section*{Appendix: Multi-Hyperuniform Disordered Point Configurations via the MultiScale Packing Model}

In this Appendix, we provide additional examples of
multi-hyperuniform disordered point configurations obtained via
the multiscale packing model for the case of three components
(red, blue and green species). These examples illustrate the
versatility and capacity of our model to generate multicomponent
systems with varying degrees of hyperuniformity (see discussion
below), apart from modeling the avian system. Specifically, we
will show that the degree of hyperuniformity of the overall
patterns can be controlled by tuning the overall final packing
fraction $\phi$ associated with the hard-core exclusion regions
for different species in the system. Note that in our model, in
the infinite-dilute limit $\phi \rightarrow 0$, i.e., in the
absence of the hard-core exclusion effects, the inherent
structures associated with the remaining {\it long-range soft-core
repulsion} are triangular lattice arrangements of points, which
are in fact the most hyperuniform point configurations in two
dimensions \cite{ref23}. As the hard-core exclusion regions for
each species grow in size (i.e., $\phi$ increases), the degree of
spatial order will be gradually reduced due to the aforementioned
geometrical frustrations caused by the hard cores, while the
system remains hyperuniform and disordered up to some packing
fraction $\phi_{C}$. Therefore, our algorithm is robust in
producing multi-hyperuniform systems with a varying degree of
disorder over a wide range of packing fractions. However, we
emphasize that there exist a threshold packing fraction
$\phi_{C}$, above which the system ceases to be hyperuniform.

In our simulations, the numbers of particles for different species
are chosen to be the same, i.e., $n_R = n_B = n_G = 500$, where
the subscripts ``R'', ``B'', ``G'' indicate the red, blue and
green species, respectively. The three species possess the same
number density and thus, the same size $R_s$ for the homotypic
repulsion [c.f. Eq. (\ref{potential})]. The relative sizes of the
hard core are respectively 1.0, 1.5 and 2.0 for red, blue and
green species. Initial configurations with an overall packing
fraction $\phi = 0.3$ are generated using the random sequential
addition process. Then the growth and relaxation procedure is
employed to generate disordered inherent structures associated
with the soft interactions at different packing fractions.

\begin{figure*}
\begin{center}
\includegraphics[width=11.5cm,keepaspectratio]{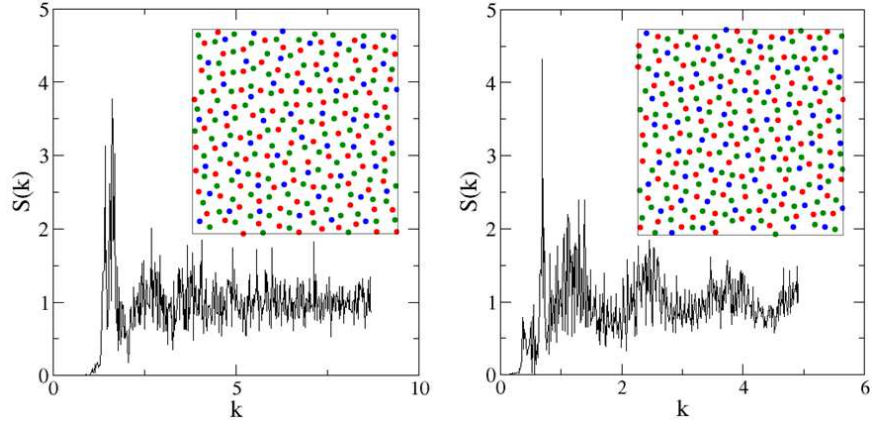} \\
\end{center}
\caption{Configurations and the associated $S(k)$ of a
three-component system for selected $\phi$ values. Left panel: At
$\phi = 0.35$, the structure factor $S(k) = 0$ for $k<K^*$ which
indicate the system is stealthy. Right panel: At $\phi = 0.55$,
the structure factor $S(k) \sim k^2$ for small $k$ values, which
indicates that number variance $\sigma^2(R) \sim R$ for large
window sizes. The system possesses different degrees of
hyperuniformity that can be ascertained from the small-$k$
behavior of $S(k)$.} \label{fig_multi}
\end{figure*}

The resulting patterns are multi-hyperuniform, i.e., both the
individual species and the overall patterns are hyperuniform.
However, here we will only focus on the degree of hyperuniformity
in the overall pattern. Fig. \ref{fig_multi} shows the
configurations and associated $S(k)$ of the \textit{overall}
system for selected $\phi$ values. At $\phi = 0.35$, the structure
factor $S(k) = 0$ for $k<K^*$, indicating that the system is
stealthy \cite{ref17} (i.e., the pattern completely suppresses
scattering of the incident radiation associated with wavenumbers
smaller than $K^*$ and thus, is transparent at the corresponding
wavelengths) and yields a higher degree of order. At $\phi =
0.55$, the structure factor is quadratic in $k$, i.e., $S(k) \sim
k^2$ for small $k$ values, which indicates that number variance
grows with the surface of the observation widow, i.e.,
$\sigma^2(R)\sim R$ for large window sizes (i.e., large $R$
values). This is to be contrasted with the large-$R$ behavior of
$\sigma^2(R)$ for the photoreceptor patterns in chicken retina,
i.e., $\sigma^2(R)\sim R \ln R$, which indicates that the number
variance grows more rapidly than that in the three-component
system associated with $\phi = 0.55$. In other words, the
three-component system at $\phi = 0.55$ possesses smaller local
number density fluctuations than those in the chicken retina,
indicating that the former is more uniform on large length scales
(i.e., displays a higher degree of hyperuniformity) than the
later.

%\end{article}

%\newpage


\begin{thebibliography}{10}
\bibitem{footnote0} We note that a more general quantity to
determine hyperuniformity in a multicomponent system is the
local-volume-fraction fluctuation as employed in Refs.
\cite{ref16, ref18, ref15, berthier}. It has been successfully
applied to ascertain hyperuniformity in maximally random jammed
packings of polydisperse hard particles, in which the
local-number-density fluctuations $\sigma^2(R)$ does not provide a
good indicator for hyperuniformity due to its lack of
particle-shape information \cite{ref16, ref15, weeks}. (Note that
in Ref. \cite{berthier}, a generalized ``compressibility''
involved in fluctuation-response relations, which is equivalent to
volume-fraction fluctuations, was used to ascertain
hyperuniformity.) However, the local-volume-fraction fluctuations
cannot be ascertained from the photoreceptor data since the
required geometrical information, such as shape and size of the
photoreceptors, are not available. The fact that the photoreceptor
patterns are effectively hyperuniform even without explicitly
considering the shape and size of the photoreceptors strongly
suggests greater suppression of infinite-wavelength
volume-fraction fluctuations could be achieved when such
information is incorporated.
\bibitem{footnote2}
The values of $\alpha$ and $\beta$ do not affect the ground state
structure. For numerical convenience, we use $\alpha = \beta = 1$
in our simulations. We note that other forms of the repulsive
interactions such as generalized Coulombic interactions $E \sim
(1/r)^\gamma$ ($\gamma\ge1$) can be used, which also lead to the
same ground state. The pair potential \eqref{potential} results in
a repulsive force along the center-to-center direction between a
pair of like-cells when their distance $r$ is smaller than $2R_s$.
The magnitude of the repulsive force is simply the derivative of
the pair potential.
\bibitem{newref43}
S. Torquato, Phys. Rev. Lett. {\bf 74}, 2156 (1995); Phys. Rev. E
{\bf 51}, 3170 (1995). In these papers, the dimensionless pressure
$p/(\rho k_B T)$ (where $\rho$ is the number density, $k_B$ is
Boltzmann's constant and $T$ is the temperature) was denoted by
$a_0$, and for $d=2$ contained a typographical error, namely,
$a_0$ should be equal to $(1-0.128\phi^2)/(1-\phi)^2$.
\bibitem{footnote}
The structure factor value at $k=0$ for identical equilibrium hard
disks can be computed from the isothermal compressibility
$\kappa_T = \frac{1}{\rho}\frac{\partial \rho}{\partial p}|_T$ of
the system via $S(0) = \rho k_B T \kappa_T$. Using the expression
for the pressure given in Ref. [43], we have $S(0) =
(1-\phi)^3/(1+\phi-0.384\phi^2+0.128\phi^3)$.
\end{thebibliography}
\end{document}